\documentclass[conference]{IEEEtran}
\IEEEoverridecommandlockouts

\usepackage{makecell}
\usepackage{booktabs}
\usepackage{multicol}
\usepackage{CJKutf8}
\usepackage{multirow} 
\usepackage{color,mathtools,setspace,comment,float,array}
\usepackage{array}
\usepackage{subcaption}
\usepackage[hidelinks, breaklinks = true]{hyperref}

\usepackage{cite}
\usepackage{amsmath,amssymb,amsfonts}
\usepackage{algorithmic}
\usepackage{graphicx}
\usepackage{textcomp}
\usepackage{xcolor}
\def\BibTeX{{\rm B\kern-.05em{\sc i\kern-.025em b}\kern-.08em
    T\kern-.1667em\lower.7ex\hbox{E}\kern-.125emX}}

\begin{document}

\title{Cell-free versus Conventional Massive MIMO : An Analysis of Channel Capacity based on Channel Measurement in the FR3 Band\\}

\author{
    Qi Zhen, Pan Tang, Haiyang Miao, Enrui Liu, Ximan Liu, Zihang Ding, Jianhua Zhang \\
    State Key Laboratory of Networking and Switching Technology \\
    Beijing University of Posts and Telecommunications, Beijing 100876, China \\
    \textit{Email: \{zq2024018002, tangpan27, hymiao, liuenrui, liuxm2020, dingzihang, jhzhang\}@bupt.edu.cn}
}

\maketitle

\begin{abstract}
Cell-free massive MIMO (CF-mMIMO) has emerged as a promising technology for next generation wireless systems, combining the benefits of distributed antenna systems (DAS) and traditional MIMO technology. In this work, we present the first extensive channel measurements for CF-mMIMO in the mid-band (FR3, 6-24 GHz), using a virtual widely distributed antenna array comprising 512 elements in the urban Macrocell (UMa) environment. Based on the measurement data, this paper compares the channel capacity of CF-mMIMO and Conventional mMIMO under both line-of-sight (LOS) and non-line-of-sight (NLOS) conditions across a range of signal-to-noise ratios (SNRs). We then analyze how channel capacity varies with Rx positions from the perspectives of the full array and of individual subarrays. Finally, we conclude that the 64-element array configuration yields the greatest advantage in channel capacity for CF-mMIMO in the measurement environment considered, with gains of 14.02\% under LOS and 24.61\% under NLOS conditions. This in-depth analysis of channel capacity in the FR3 band provides critical insights for optimizing CF-mMIMO systems in next generation wireless networks.
\end{abstract}


\section{Introduction}
The sixth generation (6G) of mobile communication needs to allocate new frequency bands to meet rising demands for connectivity, data rates, and capacity\cite{recommendation2023framework} . For 6G, the mobile communication mainly focuses on the new mid-band spectrum (FR3, 6-24 GHz) to achieve tightly integrated collaborative communication \cite{zhang2024new,Miao2023Sub-6}. However, the mobile network still suffers from the inter-cell interference and unfair services\cite{interdonato2019ubiquitous}. Due to the lack of cell boundaries in the data transmission, Cell-free massive multiple-input multiple-output (CF-mMIMO) has emerged as a promising paradigm that unifies the advantages of distributed antenna systems (DAS) and MIMO\cite{Ngo2017cell-free,interdonato2019ubiquitous}. Therefore, the combination of mid-band spectrum and CF-mMIMO technology is an effective approach to meet the high-capacity and fair-service requirements of 6G.

Signals transmit information between the transmitter (Tx) and receiver (Rx) over the wireless channel. The wireless channel critically determines the performance boundaries of the mobile communication system\cite{molisch2012wireless,zhang2020channel,zhang2023channel}. When the antenna configuration transitions from the conventional centralized array to the geographically separated subarrays, there emerges variation of channel performance during the transmission of signals. Meanwhile, compared to the Sub-6 GHz commonly used in the fifth generation (5G), the higher propagation loss and greater susceptibility to environmental factors of mid-band signals also impact the channel performance. Therefore, it is crucial to investigate the channel performance for CF-mMIMO in mid-band, and the advantage over the conventional centralized mMIMO under the same conditions.

Up to date, several studies have investigated CF-MIMO, demonstrating that it can achieve better channel performance compared with conventional centralized MIMO systems. 
\cite{Ibernon2008comparison} investigated the channel performance for conventional and distributed MIMO systems in indoor scenarios. Given that CF-MIMO is suitable for deployment in both indoor and outdoor environment, \cite{Alammari2022user-centric} conducted several simulations in dense urban scenario to evaluate how CF-mMIMO deployments outperforms co-located mMIMO in the up-link performance. To further validate the performance advantages of CF-MIMO under real-world conditions, the authors in \cite{Loschenbrand2022towards} presents a practical analysis of co-located and widely distributed antenna configurations. The authors of \cite{Simon2023measurement} conducted channel measurements in a suburban vehicular scenario. The results demonstrated superior SNR performance compared to co-located antennas. Limited studies have investigated the advantage of CF-mMIMO over conventional co-located mMIMO in terms of channel capacity in the mid-band. Moreover, the widely spaced subarrays-based architecture represents a more feasible configuration than fully distributed antenna deployment for CF-mMIMO systems  \cite{Emil2025enabling}. To the best of our knowledge, particularly limited experimental work has analyzed the channel capacity of CF-mMIMO using this antenna configuration in the mid-band. 

To address the gap in the existing literature, this paper performs a pioneering channel measurement campaign in the FR3 band for CF-mMIMO and Conventional mMIMO. \footnote{Note that ‘CF-mMIMO’ refers to systems comprising widely spaced subarrays, while ‘Conventional MIMO’ refers to co-located antenna arrays. }The novel comparative analysis of channel capacity between CF-mMIMO and Conventional mMIMO is demonstrated. The key contributions of this paper are as follows: 

\begin{itemize}

    \item An extensive channel measurement  campaign is conducted in the FR3 band. To the best of our knowledge, this work is the first channel measurement carried out at 15GHz for CF-mMIMO and Conventional mMIMO. The measurement employed a 512-element antenna array, which was virtually formed by moving the 128-element antenna array.  
    \item The comparative analysis of channel capacity versus SNR indicated that under NLOS conditions, the capacity advantage of CF-mMIMO over Conventional MIMO is more pronounced. Furthermore, the results revealed that, despite the Rx being located at the same position, the propagation conditions of signals may differ among subarrays for CF-mMIMO. 
    \item We demonstrated that the 64-element antenna configuration delivers the greatest channel capacity improvement for CF-mMIMO in this measurement environment, with the gain ratios of 14.02\% and 24.61\% under LOS and NLOS conditions, respectively.

\end{itemize}

The rest of the paper is structured as follows: Section \ref{sec:II} outlines the measurement framework and environment for Conventional mMIMO and CF-mMIMO. Section \ref{sec:IV} presents the comparative results and analysis of channel capacity between Conventional mMIMO and CF-mMIMO. Finally, Section \ref{sec:V} concludes this paper with a summary of key findings.

\section{measurement framework and environment}
\label{sec:II}

In this section, we present a set of mMIMO channel measurements conducted at 15 GHz in the UMa scenario, with the configurations of widely spaced and co-located antenna arrays.

\subsection{Measurement framework description}

\begin{figure}[htbp]
	\setlength{\abovecaptionskip}{0.3 cm}
	\centering
	\includegraphics[width=0.48\textwidth]{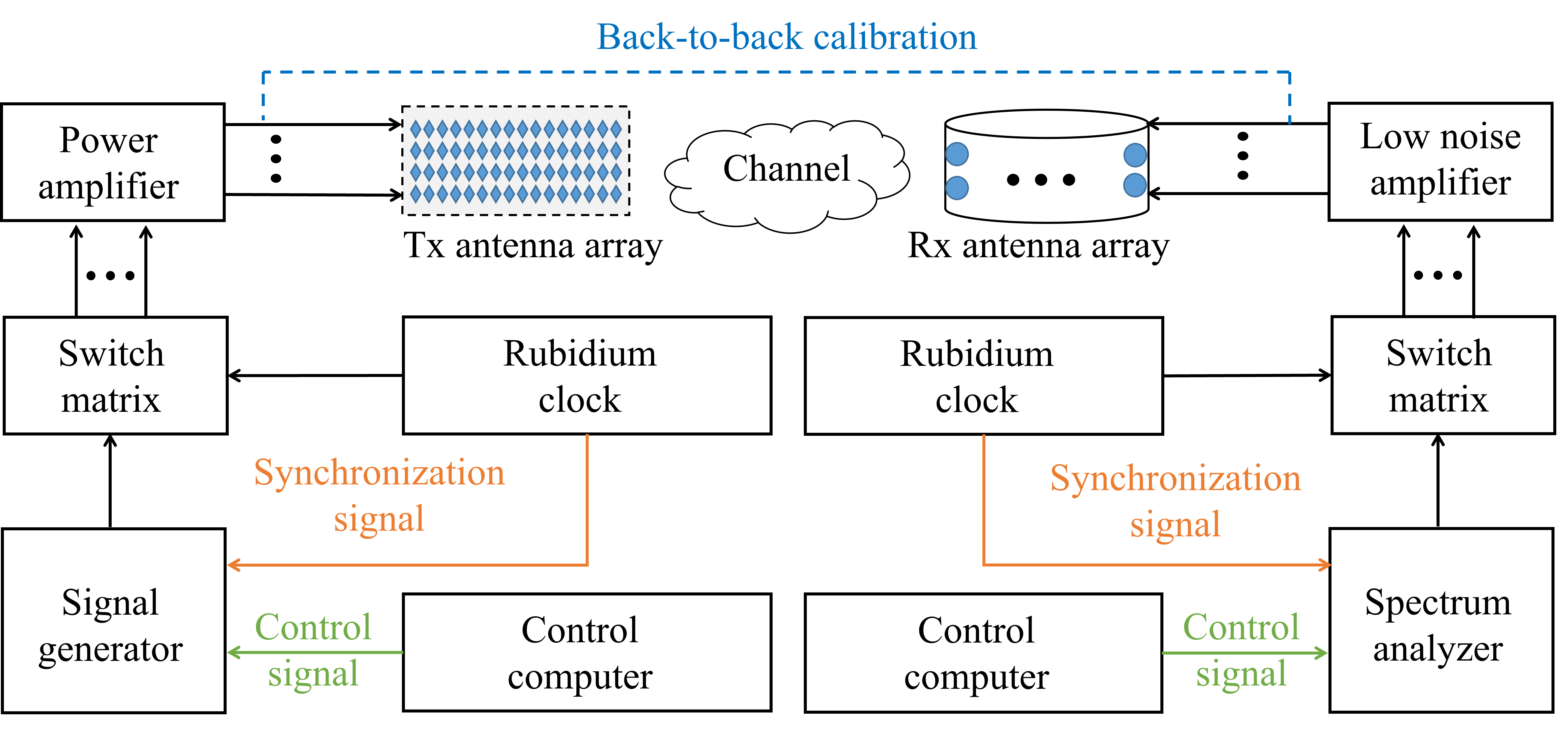}
\caption{The schematic of the channel measurement platform.}
	\label{Figure_measurement_setup}
\end{figure}

Fig.  \ref{Figure_measurement_setup} presents the measurement setup schematic. The wide-band channel measurement platform is based on time-domain correlation. The Tx generates a pseudo-random sequence and performs high-speed I/Q sampling with a signal generator and spectrum analyzer. High-precision rubidium clocks provide synchronization between the Tx and Rx, yielding nanosecond-level delay resolution. The antenna array in the Tx and Rx end uses the uniform planar antenna array (UPA) and the omnidirectional antenna array (ODA), respectively. The switch matrix enables fast switching for both the Tx and Rx\cite{Miao2025Far-field}.

\begin{figure}[htbp]
	\setlength{\abovecaptionskip}{0.3 cm}
	\centering
	\includegraphics[width=0.48\textwidth]{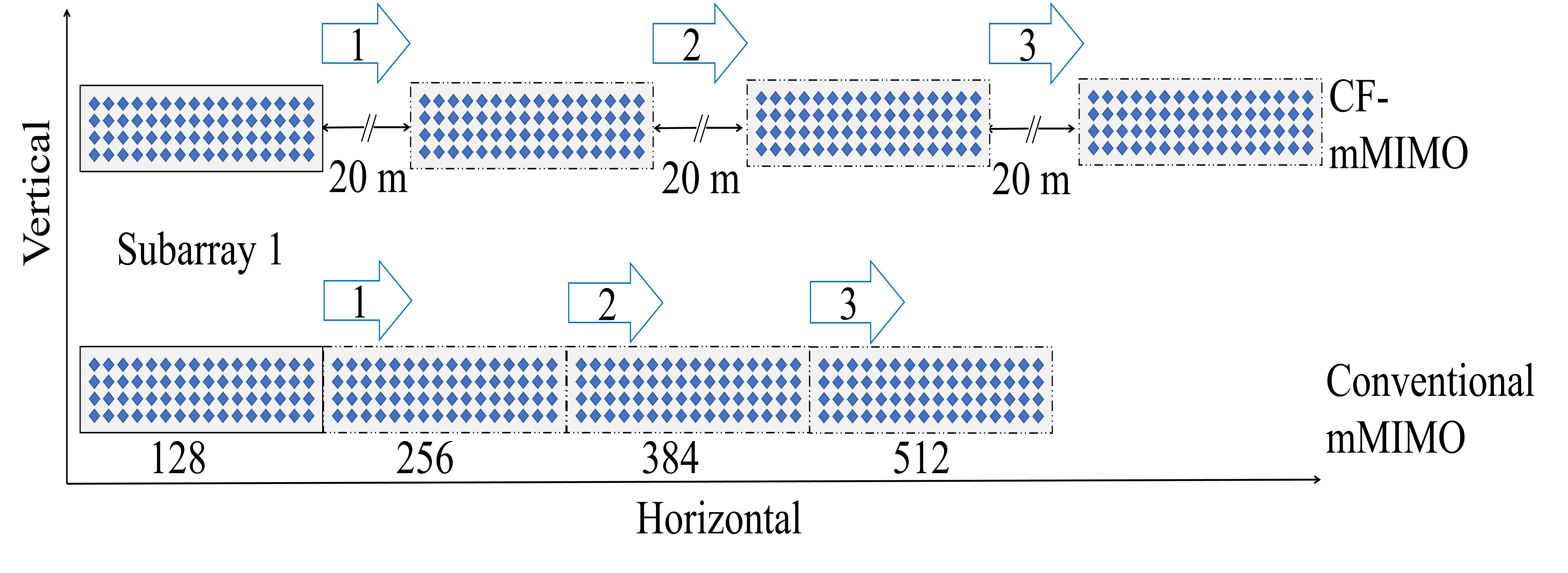}
\caption{An illustration of Tx configurations for CF-mMIMO and Conventional mMIMO.}
	\label{Figure_array_setup}
\end{figure}

For the Tx, as shown in Fig.\ref{Figure_array_setup}, there are 64 dual-polarized antennas in the panel, forming 128-element UPA. The blue diamonds in the panel represent antenna elements. The 512-element antenna array is realized through horizontal movement of the 128-element UPA. The first antenna array position is kept consistent in Conventional MIMO and CF-mMIMO configurations, as a reference for moving antenna panels. The Rx uses the 40-element ODA.

Table \ref{table_configuration_framework} lists the setup of the channel measurement framework described above. To eliminate system-induced impact and isolate the true channel impulse response (CIR), we perform back-to-back calibration prior to measurements \cite{Miao2023Sub-6}. 

\begin{table}[htbp]
	\centering
	\caption{Setup of the channel measurement framework.}
	\setlength{\tabcolsep}{0.1 mm}
	\label{table_configuration_framework}
	\renewcommand{\arraystretch}{1.5}
	 \setlength{\tabcolsep}{1 mm}
	\begin{tabular}{cc}
	 \toprule
	   \textbf{ Parameter }   & \textbf{Performance}  \\ \midrule
		Center frequency [GHz]& 14.8-15.2\\ 
		Bandwidth [MHz]& 400\\
 Tx antenna type&virtual UPA\\
 Rx antenna type& ODA\\
		Element number&   Tx : 512  /Rx : 40\\
 Tx /Rx height [m]&Tx : 27  /Rx : 1.7\\
 Antenna elements spacing [cm]&1  ($\lambda$/2)\\
 Signal modulation methods&BPSK\\
 Chip length&511\\
		Synchronous mode    & Rubidium clock\\   \bottomrule
       
	\end{tabular}
\end{table}

\subsection{Measurement environment description}

\begin{figure}[htbp]
	\setlength{\abovecaptionskip}{0.3 cm}
	\centering
	\includegraphics[width=0.4\textwidth]{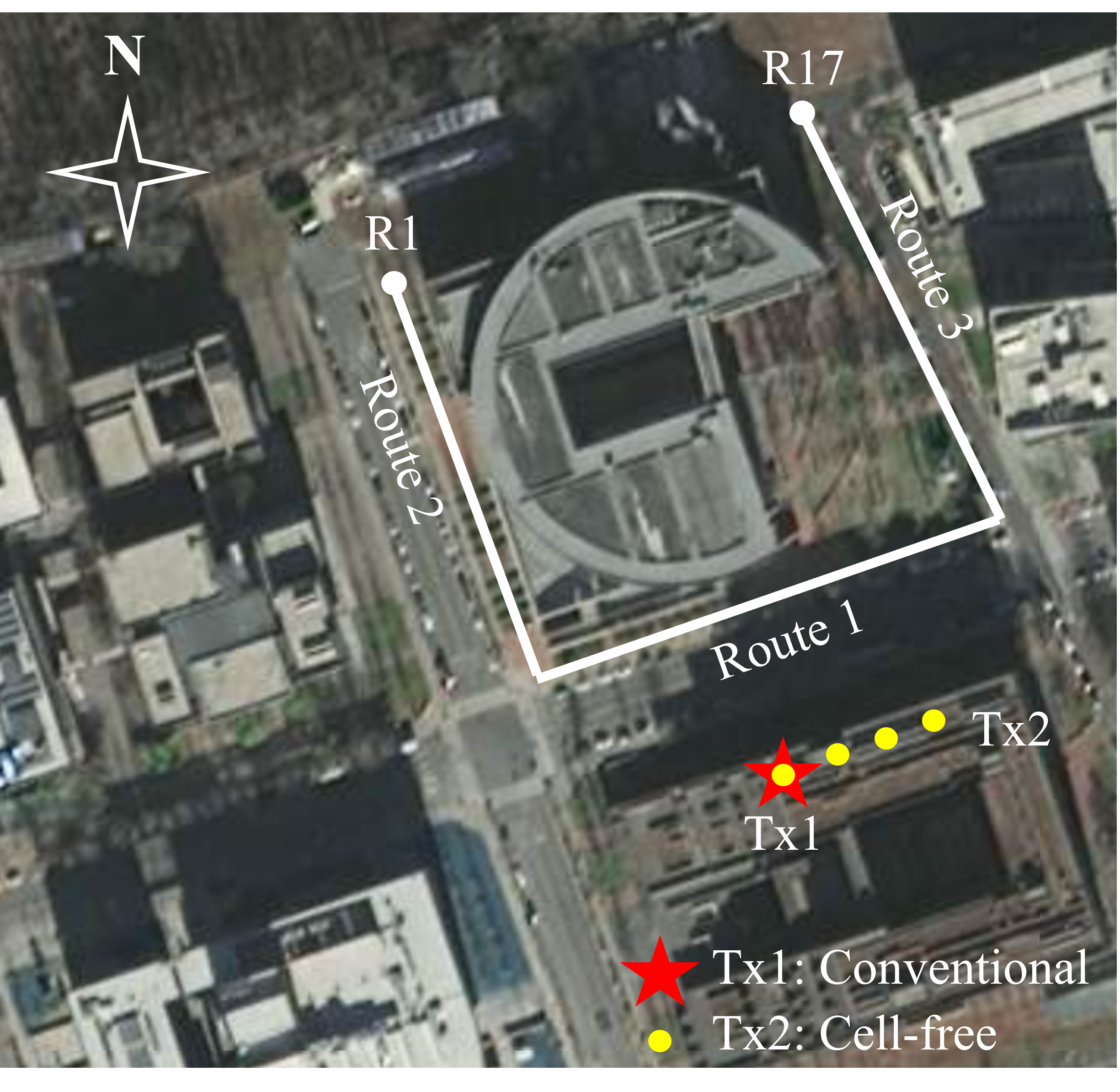}
\caption{Channel measurement environment in the UMa scenario.}
	\label{Figure_measurement_environment}
\end{figure}

We focused our channel measurements on the UMa scenario, as it represents the most frequent urban communication environment.  Fig.\ref{Figure_measurement_environment}  shows a top view of the measurement environment under consideration. The measurements were conducted in Shahe Campus of Beijing University of Posts and Telecommunications (BUPT).

\begin{figure}[h]
    \centering
    \begin{subfigure}[b]{0.24\textwidth} 
    \centering
        \includegraphics[width = \linewidth]{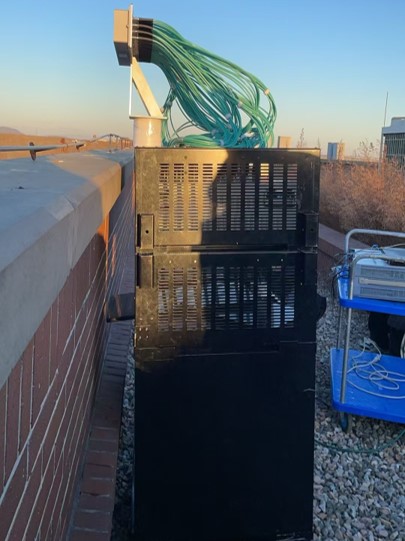}
        \caption{}
        \label{figure_Tx_real}
    \end{subfigure}
    \hfill 
    \begin{subfigure}[b]{0.24\textwidth}
    \centering
        \includegraphics[width = \linewidth]{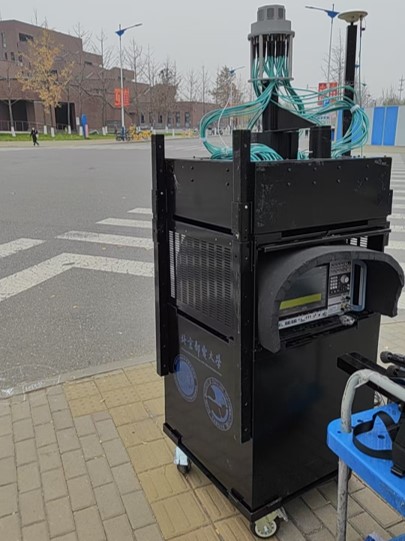}
        \caption{}
        \label{figure_Rx_real}
    \end{subfigure}
    
    \caption{The photograph of the measurement scenario  (a) The Tx. (b) The Rx. }
    \label{figure_TRx_real}
\end{figure}

The Tx is mounted on the roof top of the Academic Laboratory Complex of BUPT, at approximately 27 m above ground level, thus operating in a UMa scenario. The antenna array panel of Tx is positioned facing the library. The red star and yellow points in Fig.\ref{Figure_measurement_environment} represent the Tx location of Conventional mMIMO and CF-mMIMO, respectively. The Rx is positioned at the typical user equipment (UE) height of 1.7 meters. The ODA is fixed on a cart, which is pushed along a predefined trajectory. As shown in Fig.\ref{Figure_measurement_environment} , the trajectory is divided into three routes. Route 1 is parallel to the Tx array with the inter-point spacing of 20 m, while Routes 2 and 3 are perpendicular to Route 1, with the inter-point spacing of 24 m. The total lengths of Routes 1–3 are 149 m, 96 m, and 120 m, respectively. The library, encircled by three routes, creates signal interference for the measurement points along Route 2 and Route 3. Photographs of the Tx and Rx antennas in the actual measurement scenario are given in Fig.\ref{figure_TRx_real}. 

\section{channel capacity comparative analysis}
\label{sec:IV}
In this section, we will analyze the channel capacity versus signal-to-noise ratio (SNR), Rx positions and number of Tx elements under LOS and NLOS conditions. 

Channel capacity characterizes the channel's maximum information-carrying capability, calculated as\cite{paulraj2003introduction}: 
\begin{equation}
\begin{split}
C=\frac{1}{B} \int_B \log _2 \operatorname{det}\left(\mathbf{I}_M+\frac{\rho}{N} H(f)H^H(f)\right) d f,
\end{split} 
\end{equation}
where  $B$ is the system bandwidth, $\mathbf{I}_M$ is an $M\times M$ unit matrix, $M$ is the number of Rx antenna elements, $\rho$ is the transmission SNR, $N$ is the number of Tx antenna elements and $ (·) ^H $ is a Hermitian operation. 

When the channel state information (CSI) is unknown at the BS side, allocating equal power across all antennas is the optimal strategy for capacity calculation \cite{goldsmith2005wireless,telatar1999capacity}. The mean channel capacity then is calculated by
\begin{equation}
\begin{split}
\tilde{C}=\frac{1}{K} \sum_{k=1}^K \log _2 \operatorname{det}\left(\mathbf{I}_M+\frac{\rho}{\beta N} \mathbf{H}_r \mathbf{H}_r^H\right),
\end{split} 
\end{equation}
where $\mathbf{H}_r$ is the discrete channel realization, $K$ is the total number of such realizations. A normalization factor $\beta $ is applied to ensure uniform average channel power gain across all $\mathbf{H}_r$ in each snapshot, which can be expressed as
\begin{equation}
\begin{split}
E\left\{\frac{1}{\beta}\left\|\mathbf{H}_r\right\|_F^2\right\}=N \cdot M,
\end{split} 
\end{equation}
where $\left\|·\right\|_F$ is the Frobenius norm.

\subsection{Channel capacity versus SNR}
In this subsection, we analyze and compare the channel capacities of CF-mMIMO and Conventional mMIMO across a range of SNR values, considering both LOS and NLOS propagation conditions. 

\begin{figure}[h]
    \centering
    \begin{subfigure}[t]{0.48\textwidth} 
        \includegraphics[width=\linewidth]{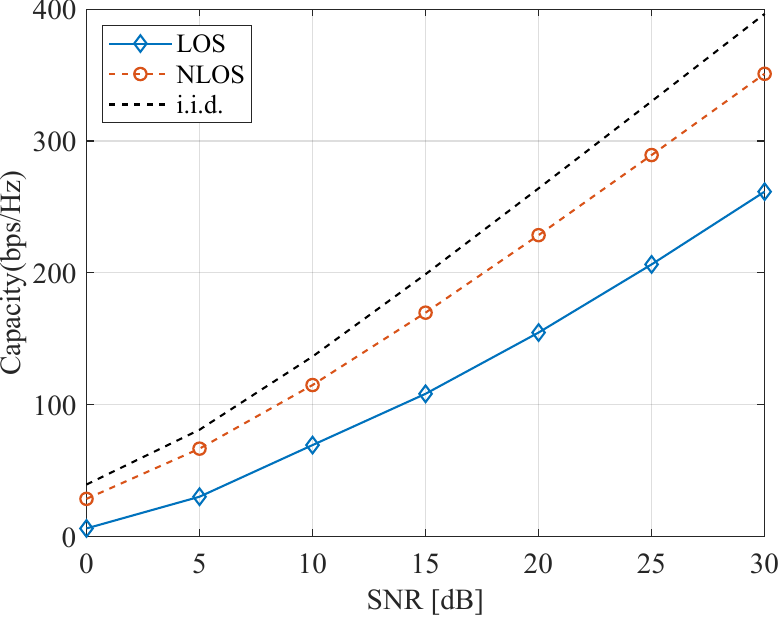}
        \caption{}
        \label{figure_capacity_CF-mMIMO}
    \end{subfigure}
    \hfill 
    \begin{subfigure}[t]{0.48\textwidth}
        \includegraphics[width=\linewidth]{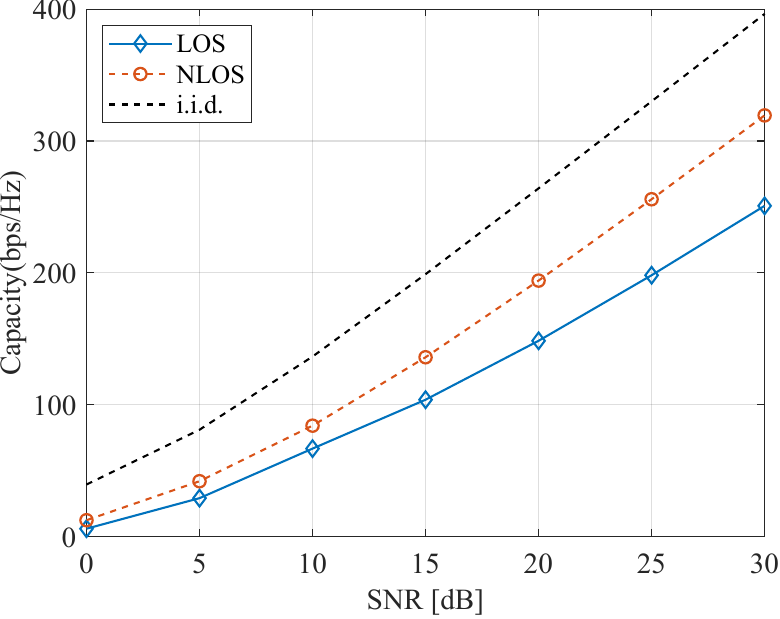}
        \caption{}
        \label{figure_capacity_Conventional_mMIMO}
    \end{subfigure}
    
    \caption{The channel capacity under different SNRs and propagation conditions  (a) CF-mMIMO. (b) Conventional mMIMO. }
    \label{figure_capacity_SNR}
\end{figure}

Fig.\ref{figure_capacity_SNR} compares the channel capacity of Conventional mMIMO with that of CF-mMIMO different SNRs and propagation conditions. It can be seen that as the SNR increases, the channel capacity exhibits a rising trend, and the channel capacity under NLOS condition is generally higher than that under LOS conditions. For a given SNR and identical signal propagation conditions, the channel capacity of CF-mMIMO consistently surpasses that of Conventional mMIMO with the same number of antenna elements. Moreover, the capacity advantage of CF-mMIMO becomes more pronounced under NLOS condition. 

\subsection{Channel capacity versus Rx positions }
In this subsection, we analyze how the channel capacity changes as the Rx moves along the measurement routes. Considering that the channel capacity differs markedly under LOS and NLOS conditions, the two cases are considered separately. 
An SNR of 25 dB—a commonly encountered value—is selected.
\begin{figure}[h]
    \centering
    \begin{subfigure}[t]{0.48\textwidth} 

        \includegraphics[width=\linewidth, clip,trim=50 60 20 60]{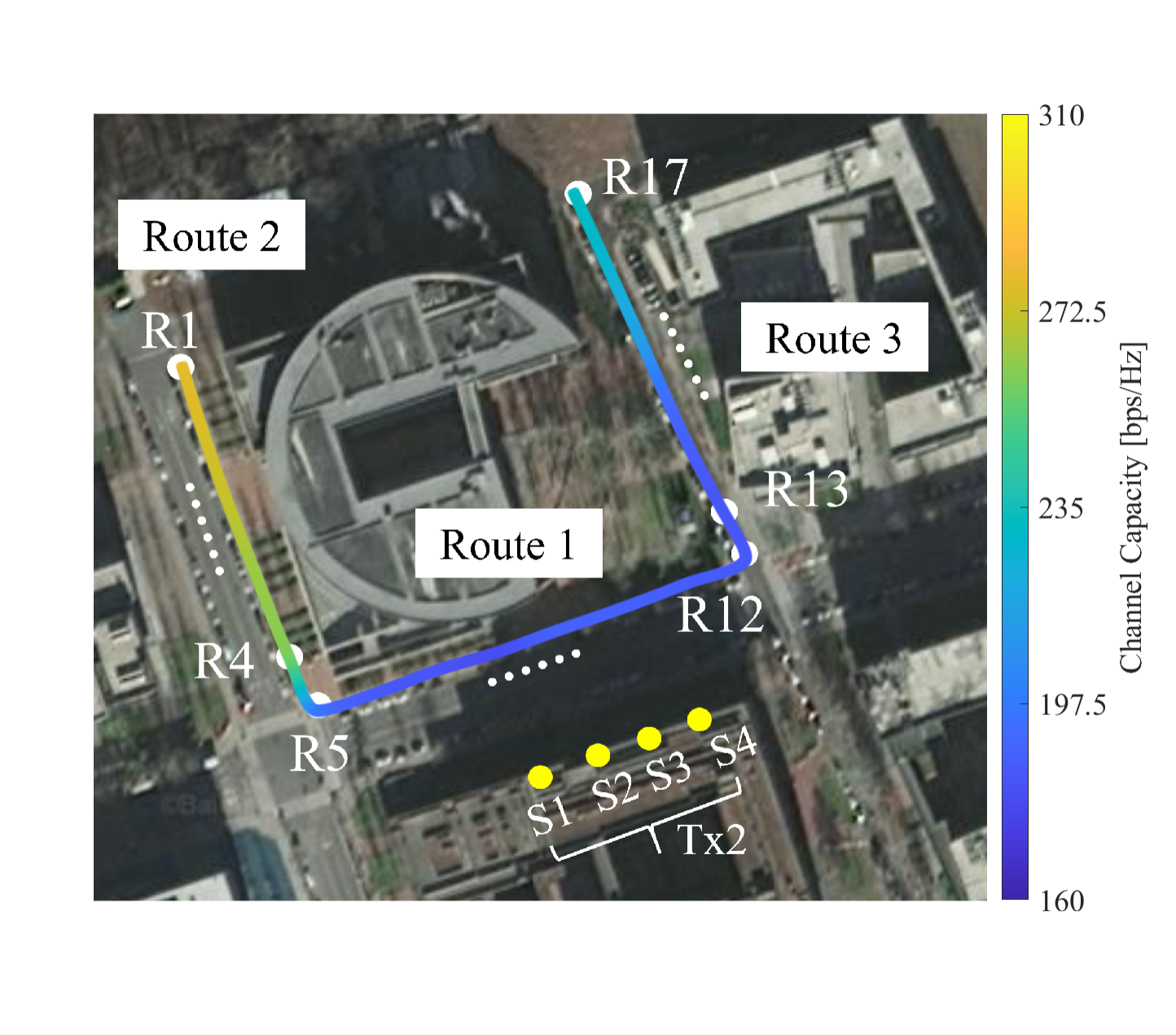}
        \caption{}
        \label{figure_capacityPoint_CF-mMIMO}
    \end{subfigure}
    \begin{subfigure}[t]{0.48\textwidth}
        \includegraphics[width=\linewidth, clip,trim=50 40 20 50]{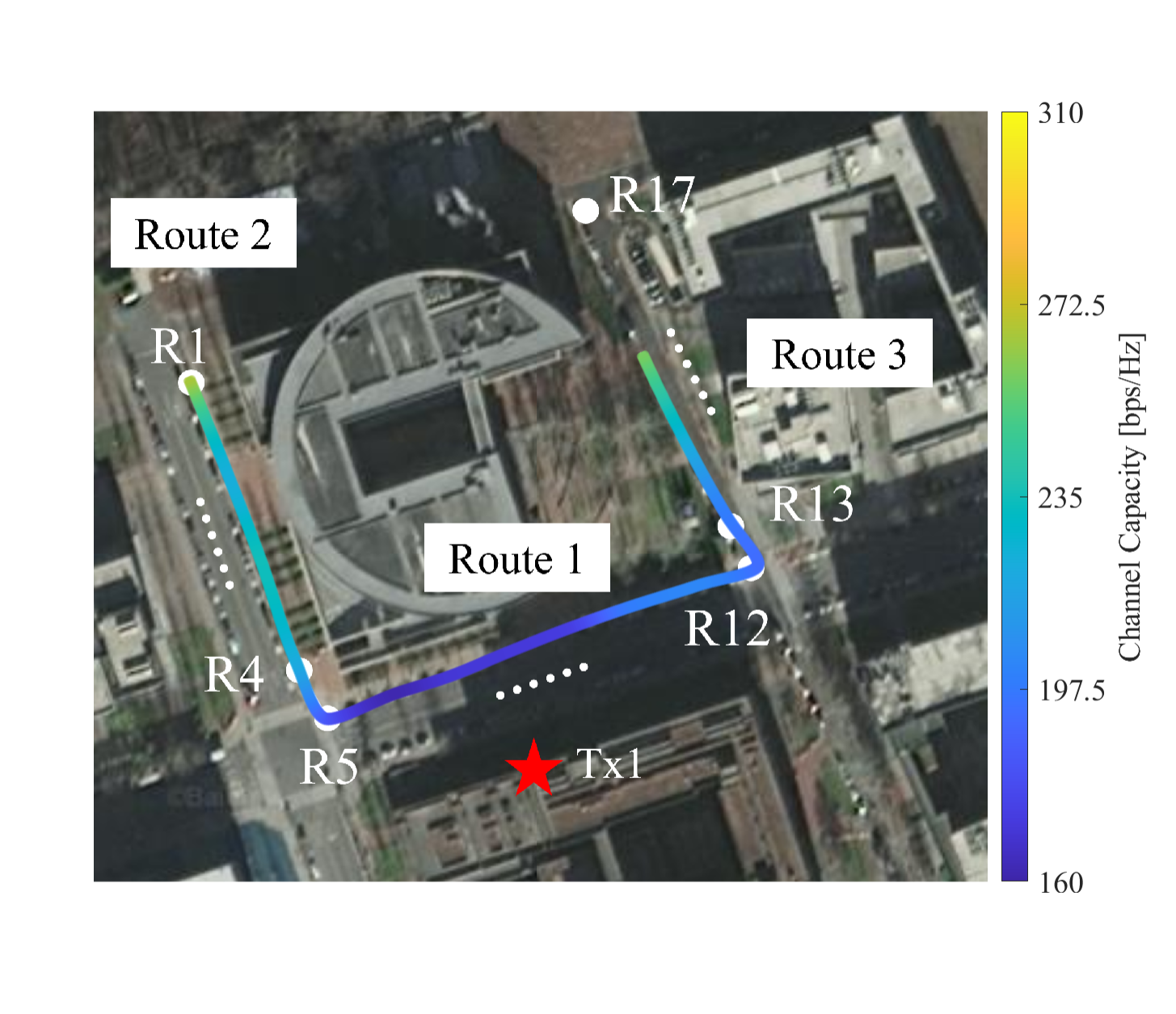}
        \caption{}
        \label{figure_capacityPoint_Conventional_mMIMO}
    \end{subfigure}
    
    \caption{Variation of channel capacity as the Rx position changes along the routes, SNR = 25 dB (a) CF-mMIMO. (b) Conventional mMIMO. }
    \label{figure_capacityPoint_SNR}
\end{figure}
Fig. \ref{figure_capacityPoint_SNR} presents heatmaps of the channel capacity variations as the Rx moves along the measurement routes, comparing the CF-mMIMO and Conventional configurations. Fig.\ref{figure_capacityPoint_Conventional_mMIMO} illustrates the Conventional mMIMO configuration, where the red pentagrams represent the virtually co-located large-scale antenna array, designated as Tx1. In Fig.\ref{figure_capacityPoint_CF-mMIMO}, the four yellow circles denote Subarrays 1 to 4, which together form a virtual large-scale antenna array referred to as Tx2. 

As illustrated in Fig. \ref{figure_capacityPoint_SNR}, at Route 2 (NLOS condition) the channel capacity of CF-mMIMO exhibits a pronounced gain of more than 35 bps/Hz compared with Conventional MIMO, whereas the enhancement at points 5-8 (LOS condition) is not very evident. This is because the four sub-arrays for CF-mMIMO, which are geographically separated, form an exceptionally large virtual antenna array. Under NLOS condition, the presence of more incomplete scatterers becomes significant for such a large-scale Tx antenna array. This, in turn, intensifies the NLOS propagation environment, leading to more dispersed multipath components (MPCs) and a substantial increase in channel capacity. 

\begin{figure}[htbp]
    \centering
    \begin{subfigure}[t]{0.48\textwidth} 
        \includegraphics[width=\linewidth]{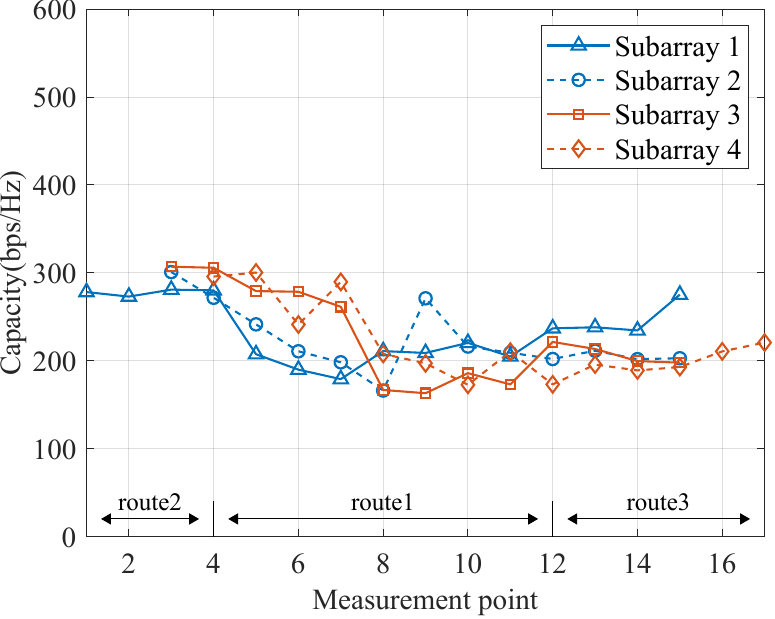}
        \caption{}
        \label{figure_capacityPointSub_CF-mMIMO}
    \end{subfigure}
    \hfill 
    \begin{subfigure}[t]{0.48\textwidth}
        \includegraphics[width=\linewidth]{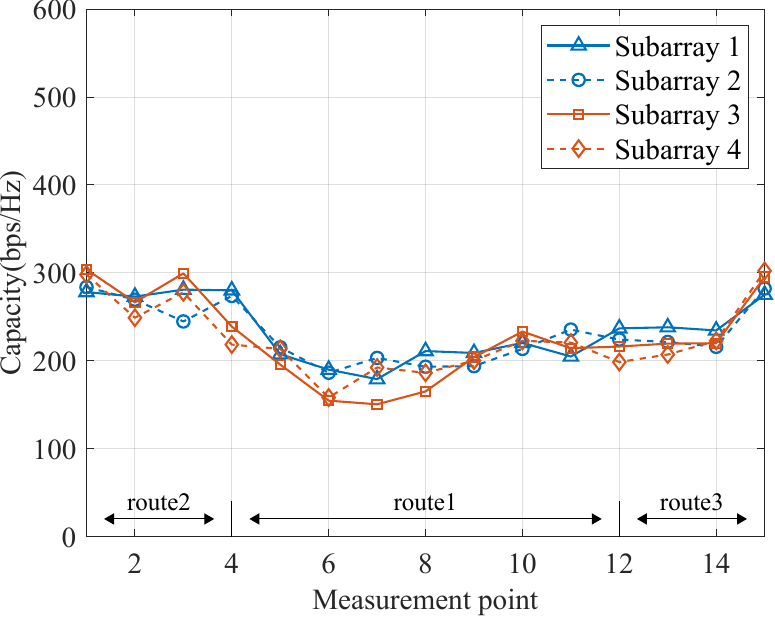}
        \caption{}
        \label{figure_capacityPointSub_Conventional_mMIMO}
    \end{subfigure}
    
    \caption{The channel capacity of the four subarrays at different Rx positions, SNR = 25 dB (a) CF-mMIMO. (b) Conventional mMIMO. }
    \label{figure_capacityPointSub_SNR}
\end{figure}

Next, the channel capacity of the four subarrays for CF-mMIMO and Conventional mMIMO at different Rx positions are analyzed, as shown in Fig.\ref{figure_capacityPointSub_SNR}. Owing to NLOS propagation condition, no signals are received by Subarray 2 to 4 at measurement points 1–2, points 16 and 17 for CF-mMIMO. Furthermore, the last detectable Conventional mMIMO signal occurs at point 15. At the farther points 16 and 17, only the CF-mMIMO configuration successfully receives signals, with channel capacity exceeding 200 bps/Hz. This result highlights the potential of CF-mMIMO to extend coverage for the next generation communication.

At certain measurement points, the Tx subarrays encounter different propagation conditions for CF-mMIMO. For example, at point 15, the channel capacity of all four subarrays in the Conventional mMIMO increases by nearly 100 bps/Hz compared with the previous point, whereas in the CF-mMIMO configuration only subarray 1 exhibits a similar abrupt rise in channel capacity. This is closely related to the antenna array placement and the signal propagation environment. For CF-mMIMO, subarrays 2–4 are positioned farther to the east compared with subarray 1, and their signals are less subject to the blockage and scattering from the opposing library building. In this way, Subarray 1 is under NLOS condition while Subarray 2-4 are under LOS condition.

\subsection{Channel capacity versus number of Tx elements }
In this subsection, we focus on examining how the number of Tx elements influences the advantage in channel capacity for CF-mMIMO over the Conventional configuration.

\begin{figure}[h]
    \centering
    \begin{subfigure}[t]{0.48\textwidth} 
        \includegraphics[width=\linewidth]{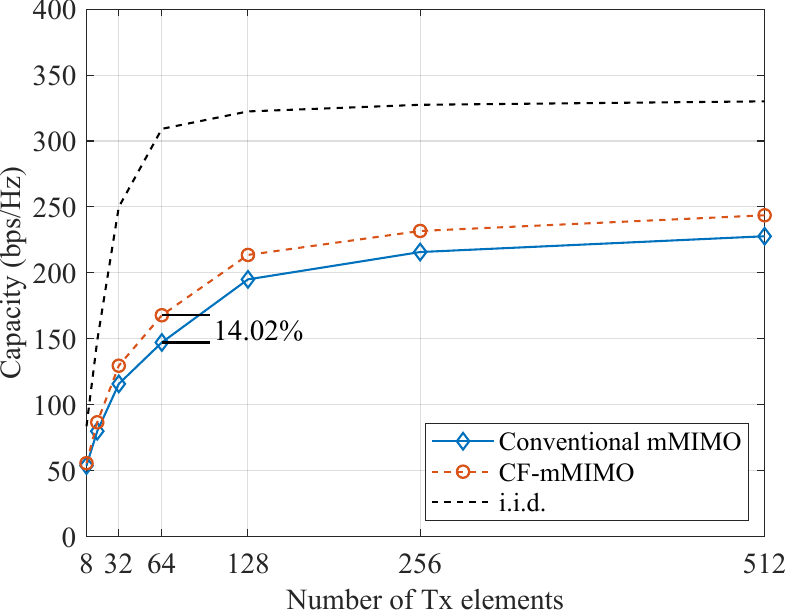}
        \caption{}
        \label{figure_capacityNTx_LOS}
    \end{subfigure}
    \hfill 
    \begin{subfigure}[t]{0.48\textwidth}
        \includegraphics[width=\linewidth]{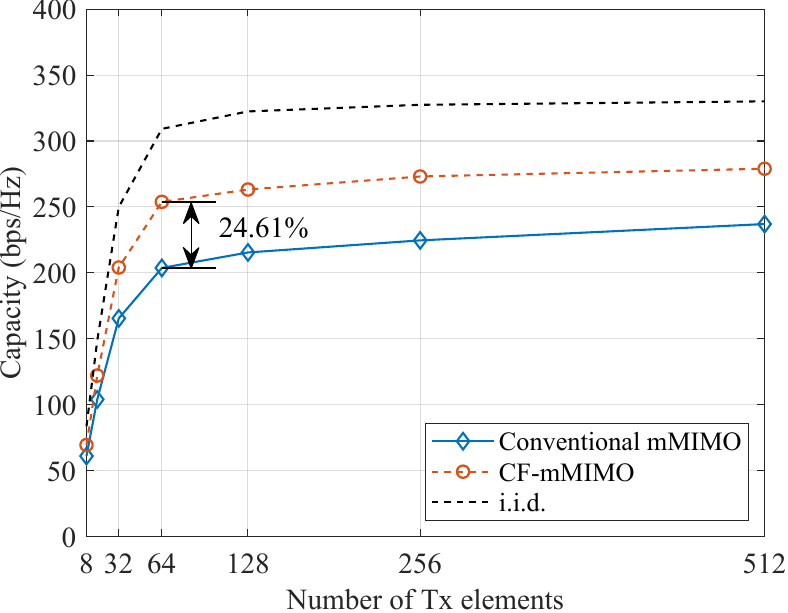}
        \caption{}
        \label{figure_capacityNTx_NLOS}
    \end{subfigure}
    
    \caption{The channel capacity of CF-mMIMO and Conventional mMIMO  under different numbers of Tx elements, SNR = 25 dB (a) LOS. (b) NLOS. }
    \label{figure_capacity_NTx}
\end{figure}

\begin{table}[!hb]
	\centering
	\caption{The channel capacity enhancement of CF-mMIMO over Conventional mMIMO under different propagation conditions(SNR = 25 dB).}
	\setlength{\tabcolsep}{0.1 mm}
	\label{table_SNR25_LOS/NLOS}
	\renewcommand{\arraystretch}{1.2}
	 \setlength{\tabcolsep}{1 mm}
	\begin{tabular}{cccc}\toprule
	 
	   \textbf{\makecell{Propagation \\condition}}& \textbf{\makecell{Antenna \\configuration}}& \textbf{\makecell{Capacity \\gain}} &\textbf{\makecell{Proportion \\with respect to \\Rayleigh channel}} \\\midrule 
		\multirow{7}{*}{LOS}& 8-element & 3.34\%&2.15\%\\
 &16-element & 8.45\%&4.54\%\\
		& 32-element & 11.91\%&5.52\%\\
 & 64-element & 14.02\%&6.67\%\\
 & 128-element & 9.51\%&5.75\%\\
 & 256-element & 7.41\%&4.88\%\\
 & 512-element & 6.98\%&4.82\%\\ \bottomrule
 \multirow{7}{*}{NLOS}& 8-element & 13.73\%&10.02\%\\
 & 16-element & 17.43\%&12.19\%\\
 & 32-element & 23.31\%&15.44\%\\
 & 64-element & 24.61\%&16.21\%\\
 & 128-element & 22.16\%&14.80\%\\
 & 256-element & 21.57\%&14.79\%\\
 & 512-element & 17.71\%&12.72\%\\\bottomrule
	\end{tabular}
\end{table}

Fig.\ref{figure_capacity_NTx} illustrates the channel capacity of CF-mMIMO and Conventional mMIMO under both LOS and NLOS conditions, for varying numbers of antenna elements in the entire Tx array. The capacity of the Rayleigh channel is also illustrated, denoted by i.i.d. Note that SNR is 25 dB. The detailed numerical values corresponding to each point in Fig.\ref{figure_capacity_NTx} are given in Table \ref{table_SNR25_LOS/NLOS}. Furthermore, the relative capacity enhancement of the CF-mMIMO architecture over the Conventional mMIMO has been quantified, and the proportion of this enhancement with respect to the Rayleigh channel has also been evaluated. 

From Fig.\ref{figure_capacity_NTx} and Table \ref{table_SNR25_LOS/NLOS}, it can be observed that, under identical propagation conditions, channel capacity increases with the growth in the number of Tx elements. The results in Table \ref{table_SNR25_LOS/NLOS} show that, under both LOS and NLOS conditions, the relative gain of CF-mMIMO over Conventional mMIMO first increases and then decreases as the total number of Tx elements grows, reaching its peak when the total number is 64. Notably, under NLOS condition, although the relative gain diminishes once the number of elements exceeds 64, it still maintains a relatively high value. For instance, with 512 elements, the channel capacity gain remains higher than that of 8- and 16-element configurations, at 17.71\%. 

\section{conclusion}
\label{sec:V}

In this paper, we have performed an extensive measurement campaign between Conventional mMIMO and CF-mMIMO in the FR3 band with 400 MHz bandwidth in UMa scenario. The measurement system is based on the virtual 512-element Tx antenna array. This work marks the first and largest CF-mMIMO channel measurement experiment carried out at 15 GHz, with Conventional mMIMO included for comparison. Based on the measurement, we observed the channel capacity versus SNR, and discovered the advantage of CF-mMIMO over Conventional mMIMO exhibit consistent trends. Subsequently, we analyzed the channel capacity versus Rx positions, and revealed that individual subarrays in the CF-mMIMO system can encounter distinct propagation conditions. Finally, from the analysis of the channel capacity versus number of Tx elements, we demonstrated that in the measurement environment considered, the 64-element antenna configuration yielded the largest improvement in channel capacity for CF-mMIMO, with gains of 14.02\% under LOS and 24.61\% under NLOS conditions. For further study, we emphasize the critical importance of integrating the newly emerged 6G communication technologies with mid-band CF-mMIMO.

\section*{Acknowledgment}

This work was supported by National Natural Science Foundation of China (62341128, 62201086, 62525101), National Key Research and Development Program of China (2023YFB2904805), Beijing Municipal Natural Fund (L243002) in part and Beijing University of Posts and Telecommunications-China
Mobile Research Institute Joint Innovation Center and BUPT-SICE Excellent Student Creative Foundation.

\bibliographystyle{IEEEtran}
\bibliography{ref}

\end{document}